\documentclass[aps]{revtex4}
\usepackage{graphicx}

\begin{document}

\title{Scattering by magnetic and spin-orbit impurities and the Josephson
current in superconductor-ferromagnet-superconductor junctions.}
\author{F.S. Bergeret$^{1}$, A.F. Volkov$^{2,3},$ and K.B.Efetov$^{2,4}$}
\affiliation{$^{(1)}$ Departamento de F\'{\i}sica Te\'{o}rica de la Materia Condensada
C-V, Universidad Aut\'{o}noma de Madrid, E-28049 Madrid, Spain\\
$^{(2)}$Theoretische Physik III,\\
Ruhr-Universit\"{a}t Bochum, D-44780 Bochum, Germany\\
$^{(3)}$Institute for Radioengineering and Electronics of Russian Academy of%
\\
Sciences,11-7 Mokhovaya str., Moscow 125009, Russia\\
$^{(4)}$L. D. Landau Institute for Theoretical Physics RAS, 119334 Moscow,
Russia}

\begin{abstract}
We analyze the Josephson current in a junction consisting of two
superconductors (S) and a ferromagnetic layer (F) for arbitrary impurity
concentration. In addition to non-magnetic impurities, we consider also
magnetic ones and spin-orbit scattering. In the limit of weak proximity
effect we solve the linearized Eilenberger equation and derive an analytical
expression for the Josephson critical current valid in a broad range of
parameters. This expression enables us to obtain not only known results in
the dirty and clean limits but also in a intermediate region of the impurity
concentration, which may be very important for comparison with experimental
data.
\end{abstract}

\maketitle


\section{Introduction}

Negative Josephson coupling has been predicted almost three decades ago \cite%
{Bul77,Footnote}. Bulaevskii et al \cite{Bul77} {obtained this coupling from
a model }of a tunnel Josephson junction containing magnetic impurities. By
tunneling from one superconductor to {the other,} electrons are scattered by
these impurities. {Explicit calculations led the authors of Ref.\cite{Bul77}
to the conclusion that} under certain conditions the Josephson critical
current $I_{c}$\ {might} change {its} sign.

Later on, Buzdin et al. suggested that a similar behavior {might} take place 
{provided} the insulating barrier {was} substituted by a ferromagnetic
metallic layer (F)\cite{Buzdin82}. Since then, the $0$-$\pi $ transition,
i.e. the change of sign of $I_{c}$, has been the focus of study of many
theoretical works (see \cite{GolubovRMP,BuzdinRMP,BVErmp} and references
therein).

However, {in spite of the theoretical progress, }the $0$-$\pi $ transition
has been observed only recently in SFS junctions with characteristic
thicknesses of the F layer of the order 100 $%
\text{\AA}%
$ or less \cite%
{Ryazanov,Kontos,Blum,Strunk,Sellier,Tsukernik,Sidorenko,Blamire}. {\
Although} a qualitative explanation for the observed drops of $I_{c}$ as a
function of thickness or temperature could be given in the framework of the
known theories, a quantitative description is, in many cases, still lacking.
This is due to the fact that the {authors of the} theoretical works
considered {either} the pure ballistic case \cite%
{Buzdin82,Valls,Petkovic,Bennemann,vecino} or the dirty limit using the
Usadel equation \cite{Buzdin91,Buzdin92}. In the pure ballistic case, the
critical current $I_{c}$ oscillates with the period $v_{F}/2h$ as a function
of the thickness $2d$ of the F layer, where $v_{F}$ is the Fermi velocity
and $h$ the exchange field in the ferromagnet. In the diffusive limit, both the
period of oscillation and the decaying length of the function $I_{c}(2d)$ 
{are equal to $\sqrt{D/2h}$, where $%
D=v_{F}^{2}\tau /3$ is the classical diffusion coefficient and $\tau $\ is
the momentum relaxation time due to the impurity scattering.

The diffusive limit is realized {provided} {either }the exchange energy $h$
is not too large or the mean free path $l$\ is short enough. More precisely,
the condition $h\tau <<1$ should be fulfilled. However, in many experiments
this is not the case. In particular, in those performed with strong
ferromagnets \cite{Blum,Tsukernik,Blamire}. Theory considering an arbitrary
value of $h\tau $ and $l$ has been presented in Ref. \cite{BVEpr01}, where
the Eilenberger equation for the condensate has been solved in the weak
proximity effect limit. In particular, it was shown that in a quasiballistic
case (or in case of strong ferromagnets), i.e., when the condition 
\begin{equation}
h\tau >>1\;  \label{QBcond}
\end{equation}%
is fulfilled, the critical current oscillates with the period $v_{F}/2h$ and
decreases exponentially over the mean free path $l=v_{F}\tau $.

In all theoretical works mentioned above the ferromagnet was modeled as a
normal metal with an exchange field acting on the spin of the conduction
electrons, while depairing factors such as spin-dependent (SD) scattering on
magnetic impurities, condensate flow (due to an internal magnetic field in
F) and spin-orbit (SO) scattering have not been taken into account, {although%
} they may play an important role in the proximity effect.

{The effect} of the spin-orbit scattering on the critical Josephson current
in SFS junctions in the diffusive limit has been studied in Ref. \cite%
{Demler,Beasley} and in a recent work \cite{Faure}, where it was shown that
the SO interaction affects the characteristic length of the decay of the
critical current with increasing the thickness of F. Also in the diffusive
limit the {effect} of the scattering on magnetic impurities was considered
in Refs. \cite{Pistolesi,Faure}.

Less attention {has been} paid though to the study of depairing effects in
the quasiballistic regime when the condition (\ref{QBcond}) is fulfilled. An
attempt to solve the Eilenberger equation with account for the SO and SD
scattering and for arbitrary $l$ was undertaken in Ref.\cite{Gusakova}.
However, the solution suggested by the authors of the latter work is valid
only in the diffusive limit. As we will see, in the limit determined by Eq.(%
\ref{QBcond}), the {expression} obtained in Ref. \cite{Gusakova} is a small
part of the total solution. The SD scattering was also considered in a
recent work \cite{Kashuba} in the case of a weak proximity effect and
arbitrary mean free path $l$. Unlike Ref.\cite{BVEpr01}, the authors
suggested an approximate solution for the Eilenberger equation and neglected
the SO scattering. Therefore, the problem of calculating the Josephson
current $I_{J}$\ in a general case of arbitrary mean free path $l$\ taking
into account different depairing mechanisms remained unsolved.

In this paper, {we attack this problem }calculating the current $I_{J}$
through an SFS structure in the general case of an arbitrary mean free path $%
l$ taking into account {diverse} depairing mechanisms. Following the method
presented in Ref.\cite{BVEpr01} we solve in the next section the Eilenberger
equation for the case of the spin dependent scattering on magnetic
impurities. As in Refs.\cite{BVEpr01,Gusakova,Kashuba}, we assume that the
proximity effect is weak, i.e., that the amplitude of {the} condensate
function induced in the ferromagnet is small. This assumption is reasonable
even for low temperatures {due to} the {large} mismatch of the electronic
parameters of the superconductor and ferromagnet {resulting in} a strong
reflection at the S/F interfaces. {Using} {the} exact solution obtained for
the condensate function we derive a general expression for the Josephson
critical current through the SFS system. This expression can be used for
calculating the current {at} arbitrary impurity concentration. In
particular, we calculate $I_{c}$ in the diffusive and the quasi-ballistic
limit. In section III we consider the {effect} of {the }SO scattering on the
critical current $I_{c}$ for these two cases.

\section{Solution for the Eilenberger equation and derivation of the
Josephson current}

We consider an SFS layered structure. The thickness of the F layer is $2d$
and the F/S interfaces are {located} at $x=\pm d$. The thickness of the S
layers is assumed to be infinite. Scattering of electrons by magnetic
impurities in a bulk superconductor has {first }been studied by Abrikosov
and Gor'kov \cite{AG} {using} microscopic Green's functions. For
non-homogeneous finite systems it is {more }convenient to use quasiclassical
Green's functions determined by the equations derived by Eilenberger, Larkin
and Ovchinnikov \cite{Eilenberger,LO}. In order to justify the applicability
of the Eilenberger equation, we also assume that the distance between {the }%
superconductors, {\it i.e.} the thickness of the F layer,  is larger than the mean
free path $l$ (see Refs. \cite{Shelankov,Zaikin}).We consider the case of a
weak proximity effect, i. e. when the amplitude of the elements of the
condensate matrix function $\hat{f}$ is assumed to be small: $|\hat{f}|<<1$.
In this case one can linearize the Eilenberger equation {that} in the
presence of {the} scattering on {non-magnetic and }magnetic impurities {takes%
} the form \cite{BVEdomWall,BuzdinRMP,BVErmp,Faure} 
\begin{equation}
\mathrm{sgn}\omega \hat{\tau}_{3}l\mathbf{e\nabla }\hat{f}_{\pm }+(\kappa 
\hat{f})_{\pm }=(1-2\lambda _{z}-\lambda _{\perp })\langle \hat{f}_{\pm
}\rangle +\lambda _{\perp }\langle \hat{f}_{\mp }\rangle \;  \label{Eil}
\end{equation}%
where $\hat{\tau}_{3}$ is the $z$-component of the Pauli matrices $\mathbf{%
\hat{\tau}},\mathbf{e=v}_{F}/v_{F}$ is unit vector in the direction of the
Fermi velocity, $\kappa _{\pm }=1+2(|\omega |\pm ih_{\omega })\tau _{t},$ $%
\omega \equiv \omega _{n}=\pi T(2n+1)$ is the Matsubara frequency $h_{\omega
}=h\mathrm{sgn}\omega $, $\tau _{t}^{-1}=$ $\tau _{N}^{-1}+\tau _{Mt}^{-1}${%
\ is the total scattering rate and }$\tau _{N}${\ is the momentum relaxation
time due to scattering by nonmagnetic impurities. The rate }$\tau
_{Mt}^{-1}= ${\ }$\tau _{M}^{-1}/(1+\alpha _{z}+\alpha _{\perp })${\ is the
total scattering rate due to scattering by magnetic impurities. The
parameters }$\alpha _{z},\alpha _{\perp }${\ characterize the spin-dependent
scattering. The potential of interaction with magnetic impurities can be
written in the form} 
\[
U(\mathbf{r})=U_{o}(\mathbf{r})+U_{S}(\mathbf{r})\mathbf{S\cdot \sigma }/S, 
\]
{where the potential }$U_{o}(r)${\ describes interaction with a
spin-independent part of magnetic impurities. The impurity spin S is assumed
to be classical. The coefficients }$\alpha _{z}${\ and }$\alpha _{\perp }${\
are given by }$\alpha _{z}=[|U_{S}|^{2}/|U_{o}|^{2}]\langle S_{z}^{2}\rangle
/S^{2}${\ and }$\alpha _{\perp }=[|U_{S}|^{2}/|U_{o}|^{2}]\langle
S_{x}^{2}\rangle /S^{2}+[|U_{S}|^{2}/|U_{o}|^{2}]\langle S_{y}^{2}\rangle
/S^{2}$ {. These coefficients are related to a spin-dependent scattering
rate }$\tau _{m}^{-1}${\ used in Ref. \cite{Faure}. For example, }$\alpha
_{z}=2(\tau _{M}/\tau _{m})\langle S_{z}^{2}\rangle /S^{2}${. }The angle
brackets denote averaging over angles. {The coefficients }$\lambda
_{z},\lambda _{\perp }${\ are defined as: }$\lambda _{z}=(\alpha _{z}/\tau
_{Mt})/(\tau _{Mt}^{-1}+\tau _{N}^{-1})${\ and }$\lambda _{\perp }=(\alpha
_{\perp }/\tau _{Mt})/(\tau _{Mt}^{-1}+\tau _{N}^{-1})${. Note that }$\alpha
_{z,\perp }<<1${\ \cite{AG}.}

In the general case the condensate function $\check{f}$ is a $4\times 4$
matrix in the particle-hole (Gor'kov-Nambu)$\otimes $spin space. However, in
the case of a homogeneous magnetization {considered here} and in the absence
of the spin-orbit scattering (the SO scattering will be taken into account
in the next section), {the function} $\check{f}$ is diagonal in the spin
space. Then, the function $\hat{f}_{+}$ in Eq. (\ref{Eil}) is defined as $%
\hat{f}_{+}=\check{f}_{\alpha \alpha }$ with $\alpha =1$ ($\alpha $ is the
spin index), while the other diagonal element is given by $\hat{f}_{-}=%
\check{f}_{22}=$ $-\hat{f}_{+}(-h)$. In the present case of a planar
geometry, the function $\hat{f}$ depends only on the coordinate $x$.

It is convenient to represent $\hat{f}$ as a sum of a symmetric and
antisymmetric part with respect to the momentum direction: $\hat{f}_{\pm
}(x)=\hat{s}_{\pm }(x)+\hat{a}_{\pm }(x)$. As follows from Eq.(\ref{Eil}),
the antisymmetric part $\hat{a}(x)$ is related to the symmetric one by the
expression 
\begin{equation}
\hat{a}_{\pm }=-\mathrm{sgn}\omega \left( \mu l/\kappa _{\pm }\right) \hat{%
\tau}_{3}\partial \hat{s}_{\pm }/\partial x  \label{Antisym}
\end{equation}%
while the symmetric part $\hat{s}(x)$ in the ferromagnetic region ($-d<x<d$)
obeys the equation 
\begin{equation}
\mu ^{2}l^{2}\partial ^{2}\hat{s}_{\pm }/\partial x^{2}-\kappa _{\pm }^{2}%
\hat{s}_{\pm }=-\kappa _{\pm }(h)\left[ (1-2\lambda _{z}-\lambda _{\perp
})\langle \hat{s}_{\pm }\rangle +\lambda _{\perp }\langle \hat{s}_{\mp
}\rangle \right] \;  \label{EqSym}
\end{equation}%
where $l=v_{F}\tau _{t}$ is the mean free path and $\mu =v_{x}/v_{F}=\cos
\theta $. These two equations {should be} complemented by the boundary
condition \cite{Zaitsev} 
\begin{equation}
\hat{a}\mid _{x=\pm d}=\mp \gamma (\mu )\mathrm{sgn}\omega \left( \hat{\tau}%
_{3}\hat{f}_{s}\right) \;,\mu >0  \label{BC}
\end{equation}%
where $\gamma (\mu )=T(\mu )/4$ and $T(\mu )$ is the transmission
coefficient. The latter is assumed to be small and therefore the matrix $%
\hat{f}_{s}$ in the superconductor has its bulk form $\hat{f}_{s}(\pm
d)=f_{s}(\hat{\tau}_{2}\cos (\varphi /2)\pm \hat{\tau}_{1}\sin (\varphi /2))$%
, where $f_{s}=\Delta /\sqrt{\omega ^{2}+\Delta ^{2}}$. Note that the
boundary condition Eq. (\ref{BC}) does not take into account spin-flip
processes at the interface. Boundary conditions for magnetically active
interfaces were derived in Ref. \cite{millis} and used in Ref. \cite{eschrig}
by calculating the supercurrent through a pure ballistic half-metallic layer.

In order to derive an expression for the Josephson critical current avoiding
straightforward but cumbersome calculations we first neglect the terms
proportional to $\lambda _{\perp }$. This term will be included again at the
end of this section where we will present the final expression for the
current. The way how to obtain the solution of Equations (\ref{EqSym}-\ref%
{BC}) was presented in Ref. \cite{BVEpr01}. One formally extends the
solution over the whole $x$-axis, performs a Fourier transformation and
obtains finally the transformed function $\hat{s}_{k\pm }$ from an algebraic
equation. Following this scheme we obtain an expression describing the
spatial dependence of the condensate function { 
\begin{equation}
\hat{s}_{\pm }(x)=\pm \int \frac{dk}{2\pi }\frac{2\kappa _{\pm }l}{M_{k\pm
}(\mu )}\sum_{n=-\infty }^{\infty }\hat{f}_{s,n}\exp (ikd(2n+1)-ikx)\left[
\kappa _{\pm }(1-2\lambda _{z})N_{\pm }^{-1}\left\langle \frac{\mu \gamma
(\mu )}{M_{k\pm }(\mu )}\right\rangle +\mu \gamma \right] \;\;  \label{S(x)}
\end{equation}%
with $\hat{f}_{s,n}=f_{s}[\hat{\tau}_{2}\cos (\varphi /2)+(-1)^{n}\hat{\tau}%
_{1}\sin (\varphi /2)]$, $M_{k\pm }(\mu )=(k\mu l)^{2}+\kappa _{\pm }^{2}$
and $N_{\pm }=1-(1-2\lambda _{z})\kappa _{\pm }\langle M_{k\pm }^{-1}(\mu
)\rangle $ . }

The average $\kappa \langle M_{k}^{-1}(\mu )\rangle $ can easily be
calculated (we drop the subindex $\pm $){\ 
\begin{equation}
\kappa \langle M_{k}^{-1}(\mu )\rangle =\frac{1}{kl}\tan ^{-1}\frac{kl}{%
\kappa }=\frac{1}{2ikl}\ln (\frac{i\kappa -kl}{i\kappa +kl})  \label{Average}
\end{equation}%
} The $x$-dependence of the condensate function, in particular, its
exponential decay is determined by the singular points of the integrand in
the complex $k$ plane of Eq. (\ref{S(x)}). From Eq. (\ref{Average}) one can
immediately see, that the function in the integrand has poles at $kl=\pm
i\kappa /|\mu |$ and branch points at $kl=\pm i\kappa _{\pm }$. The poles
lead to the exponential decay of $\hat{s}(x)$ like $s_{pol}(x)\sim
C_{pol}(\mu )\exp (-\kappa (d+x)/l|\mu |)$ (in the vicinity of the left
superconductor), whereas the branch points lead to terms in the solution {%
that} decay as $s_{br}(x)\sim C_{br}(\mu )\exp (-\kappa (d+x)/l)$ (where $%
-d<x$)$.$ This means that in the limit $T\tau <1$ the terms $s_{pol}(x)$
exponentially decrease over an angle-dependent distance of the order of the
mean free path $l$. On the other hand, the terms $s_{br}(x)$ decay
exponentially over an angle-independent distance of the order $l$. Both {the}
terms, $s_{pol}(x)$ and $s_{br}(x)$, oscillate with the periods $h/v_{F}|\mu
|$ and $h/v_{F}$, respectively. The amplitudes $C_{pol}(\mu )$ and $%
C_{br}(\mu )$ depend on the parameters of the system.

In the general case their calculation is rather complicated and a numerical
analysis is needed. However, as we will see in the next paragraphs, in the
quasi-ballistic and diffusive cases one can obtain explicit expressions for
the condensate function $\hat{s}(x)$. If one is interested in an
intermediate case it is convenient for numerical calculations to perform the
integration over momentum in Eq. (\ref{S(x)}) {taking into account
the relation (see, for example \cite{Abrikosov})}

{
\begin{equation}
\sum_{n=-\infty }^{\infty }\exp (i2nkd)=\frac{\pi }{d}\sum_{m=-\infty
}^{\infty }\delta (k-k_{m}),\;\;
\end{equation}%
where } $k_{m}=\pi m/d.$

Writing the condensate matrix as $\hat{s}_{\pm }=s_{1}^{\pm }\hat{\tau}%
_{1}+s_{2}^{\pm }\hat{\tau}_{2}$ one finally obtains{\ 
\begin{eqnarray}
s_{1}^{\pm } &=&\pm \frac{l\kappa _{\pm }}{d}\sum_{m=-\infty }^{\infty }%
\frac{\exp [i(x/d-1)(2m+1)\pi /2]}{M_{2m+1}^{\pm }}\left[ \kappa _{\pm
}(1-2\lambda _{z})(N_{2m+1}^{\pm })^{-1}\left\langle \frac{\mu \gamma (\mu )%
}{M_{2m+1}^{\pm }(\mu )}\right\rangle +\mu \gamma \right] f_{s}\sin \varphi
/2 \\
s_{2}^{\pm } &=&\pm \frac{l\kappa _{\pm }}{d}\sum_{m=-\infty }^{\infty }%
\frac{\exp [i(x/d-1)m\pi ]}{M_{2m}^{\pm }}\left[ \kappa _{\pm }(1-2\lambda
_{z})(N_{2m}^{\pm })^{-1}\left\langle \frac{\mu \gamma (\mu )}{M_{2m}^{\pm
}(\mu )}\right\rangle +\mu \gamma \right] f_{s}\cos \varphi /2
\end{eqnarray}%
where $M_{m}^{\pm }=(\mu l\pi m/2d)^{2}+\kappa _{\pm }^{2}$ and $N_{m}^{\pm
}=1-(1-2\lambda _{z})\kappa _{\pm }\langle 1/M_{m}^{\pm }\rangle $ }.
Knowing the condensate function induced in the ferromagnetic region one can
calculate the Josephson dc current $I_{J}$ through the SFS junction. This
current is given by 
\begin{equation}
I=(\pi iT/4)(e^{2}/\hbar )(k_{F}^{2}S/\pi ^{2})Tr\left[ \hat{\tau}%
_{3}\otimes \hat{\sigma}_{0}\sum_{\omega =-\infty }^{\infty }\left\langle
\mu \gamma \lbrack \check{s}(d),\check{f}_{s}(d)]_{-}\right\rangle \right] \;
\label{JosCur}
\end{equation}%
where  $[\check{s},%
\check{f}_{s}]_{-}=\check{s}\check{f}_{s}-\check{f}_{s}\check{s}$, $\sigma
_{0}$ is the unit matrix and {the symbol }$Tr$ {stands for the trace over
the }$4\times 4$ matrices (see \cite{BVEpr01}).

Using Eq.(\ref{S(x)}), one can write the current in the form $I=I_{c}\sin
\varphi $ with 
\begin{equation}
I_{c}=A(2\pi T)\mathrm{Re}\sum_{m=-\infty,\omega \geq 0}^{\infty
}f_{s}^{2}(-1)^{m}\langle \mu \gamma (\mu )\frac{l}{d}\frac{\kappa^+ }{
M^+_{m}(\mu )}[\frac{\kappa^+ }{N^+_{m}}(1-2\lambda _{z})\left\langle \frac{\mu
^{\prime }\gamma (\mu ^{\prime })}{M^+_{m}(\mu ^{\prime })}\right\rangle _{\mu
^{\prime }}+\mu \gamma (\mu )]\;\rangle _{\mu }\;  \label{I_c}
\end{equation}%
where $A=((e^{2}/\hbar )(k_{F}^{2}S/\pi ^{2})$ and $M_{m}(\mu )\equiv
M_{m}^{+}(\mu )$. Eq.(\ref{I_c}) {is the most general expression for the
Josephson current in terms of the solution of the Eilenberger equation}. In
the next sections we give expressions for the critical current $I_{c}$ in
the quasiballistic and diffusive limits.

\subsection*{The quasiballistic case: $|\protect\kappa |>>1$}

The {quasiballistic} case corresponds either to a strong ferromagnet ($%
h>\tau ^{-1}$) or to a clean sample ($T\tau >1$). As follows from Eq.(\ref%
{Average}), in this case $N\approx 1$ and the second term in the square
brackets in Eq.(\ref{S(x)}) is much larger than the first one. Calculating
the residue at the pole $kl=i\kappa /|\mu |,$ we obtain for $\hat{s}%
(x)\equiv \hat{s}_{+}(x)$ 
\begin{equation}
\hat{s}(x)=\gamma (\mu )f_{s}\left[ \hat{\tau}_{2}\frac{\cosh (x/L_{qb}(\mu
))}{\sinh (d/L_{qb}(\mu ))}\cos (\varphi /2)+\hat{\tau}_{1}\frac{\sinh
(x/L_{qb}(\mu ))}{\cosh (d/L_{qb}(\mu ))}\sin (\varphi /2)\right]
\label{S(x)bal}
\end{equation}%
with the length $L_{qb}(\mu )$ characterizing the quasi-ballistic case
defined as $L_{qb}(\mu )=l|\mu |/\kappa $.

It is seen for example that in the vicinity of the left superconductor,
i.e., at $1<<(x+d)/l<<d/l$ the condensate function $\hat{s}(x)$ oscillates
with the period $\pi v_{F}/h|\mu |$ decaying over the angle-dependent mean
free path $l|\mu |:s(x)\sim \exp (-kx/l|\mu |)$ \cite{BVEpr01}. Thus, if the
exchange energy is large, $h>\tau ^{-1}$, the function $\hat{s}(x)$
experiences many oscillations over the decay length $l$. Note that the
period of oscillations and the decay length strongly depend on the angle $%
\mu =\cos \theta $.

This result contradicts the conclusion of Ref.\cite{Gusakova} where the
solution for Eq.(\ref{EqSym}) {was taken in a form of an} exponential
function with an angle-independent exponent. As can be understood from Eq.(%
\ref{S(x)}), the solution obtained in Ref. \cite{Gusakova} is not general
and corresponds only to a contribution from the branch points, i.e., from
the first term in the square brackets in Eq.(\ref{EqSym}). However, in the
limit of large $\kappa $ this term is small compared to the other one (see %
\cite{VBEcom}).

We see {from Eq. (\ref{EqSym})} that in the quasiballistic case the
depairing leads only to a renormalization of the scattering time $\tau
\rightarrow \tau _{t}$.

The main contribution to the critical Josephson current {in the
quasiballistic regime} stems from the second term in the square brackets in
Eq. (\ref{I_c}). Calculating the residues at the poles of $\kappa
_{k}^{2}(\mu ^{\prime })$, we obtain (\textit{cf.} Ref. \cite{BVEpr01}) 
\begin{equation}
I_{c}=A(2\pi T)\sum_{\omega \geq 0}^{\infty }f_{s}^{2}\mathrm{Re}%
\left\langle \frac{\mu ^{\prime }\gamma ^{2}(\mu ^{\prime })}{\sinh
(2d/L_{qb}(\mu ^{^{\prime }}))}\right\rangle _{\mu ^{\prime }}\;
\label{I_cQB}
\end{equation}%
If the thickness of the F layer {considerably} exceeds the value of $%
L_{qb}(1)$, one obtains for the {critical current} $I_{c}$ 
\begin{equation}
I_{c}=A(2\pi T)\sum_{\omega \geq 0}^{\infty }f_{s}^{2}\gamma ^{2}(1)\frac{%
\sin (4hd/v_{F})}{4hd/v_{F}}\;\exp \left[ -(2d/l)(1+2\omega \tau _{t})\right]
\label{I_cQBlim}
\end{equation}

Eqs. \ (\ref{I_cQB}-\ref{I_cQBlim}) fit well {recent }experimental data \cite%
{Blum,Tsukernik}

\subsection*{The diffusive case: $|\protect\kappa |\sim 1$.}

In {the diffusive} limit the conditions $h\tau $, $T\tau <<1$ hold and the
characteristic length of the spatial variation of $\hat{s}(x)$ is much
larger than the mean free path $l$. Therefore, {we obtain from} {the
following expressions for the parameters:} $N_{k}=\kappa
_{dif}^{2}+(1-2\lambda _{z})(kl)^{2}/3$ and $k^{2}\approx M_{k}(\mu )\approx
1$ (\textit{cf} Eq. (\ref{Average})).

The behavior of $\hat{s}(x)$ at distances from the superconductors larger
than the mean free path $l$ is determined by the residue {of} the pole in $%
N_{k}$. Equation (\ref{S(x)}) finally yields 
\begin{equation}
\hat{s}(x)=\frac{\sqrt{3(1-\lambda _{z})}}{\kappa _{dif}}\langle \gamma (\mu
)\mu \rangle f_{s}\left[ \hat{\tau}_{2}\frac{\cosh (x/L_{dif})}{\sinh
(d/L_{dif})}\cos (\varphi /2)+\hat{\tau}_{1}\frac{\sinh (x/L_{dif})}{\cosh
(d/L_{dif})}\sin (\varphi /2)\right]  \label{SolSymDif}
\end{equation}%
where $\kappa _{dif}^{2}=\lambda _{z}+2(|\omega |-ih_{\omega })\tau _{t},$
and the characteristic length $L_{dif}$ of the condensate {decay} in F is $%
L_{dif}=l\sqrt{(1-2\lambda _{z})/3\kappa _{dif}^{2}}\approx l/(\kappa _{dif}%
\sqrt{3}).$

If the exchange energy $h$ is much smaller than the spin-dependent
scattering rate, $h\tau _{t}<<\lambda _{z}$, we obtain for the
characteristic length $L_{dif}\approx l/\sqrt{3\lambda _{z}}=\sqrt{D\tau
_{sp}/2},$ where $\tau _{sp}\equiv \tau /\lambda _{z}$. Thus, as expected,
the decay length in this case is related to the spin-dependent relaxation
time (see, for example, \cite{Deutscher}). In the opposite limit, $h>>1/\tau
_{sp}$, the characteristic length $L_{dif}$ is given by the well known
expression \cite{Buzdin91,BuzdinRMP,Pistolesi,Faure}: $L_{dif}\approx
l/[(1-i)\sqrt{3h\tau _{t}}]=(1/2)(1+i)\sqrt{D/h}$. Note that in this case
the condensate function both decays and oscillates on the same length $\sqrt{%
D/h}.$

If one takes into account an internal magnetic field $B$ inside the
ferromagnet given by $B=4\pi M$, the length $L_{dif}$ in the limit $h\tau
_{t}<<\lambda _{z}$ is equal to: $L_{dif}=1/\sqrt{(2dB/\phi _{0})^{2}+(D\tau
_{sp}/2)^{-1}}$, where $M$ is the magnetization in F and $\phi _{0}=\pi
\hbar c/e$ is the magnetic flux quantum (see, e.g., \cite{VZK}).

In the diffusive limit, the critical Josephson current $I_{c}$ is determined
by the first term in the square brackets of Eq. (\ref{I_c}), i.e., by the
poles of the function $N_{k}^{-1}$. Then, we find 
\begin{equation}
I_{c}=A(2\pi T)\sum_{\omega \geq 0}^{\infty }\sqrt{3}f_{s}^{2}\mathrm{Re}%
\left[ \frac{1}{\kappa _{dif}^{2}}\frac{1}{\sinh (2d/L_{dif})}\right] \;
\label{I_cDIF}
\end{equation}%
where $L_{dif}$ is defined in Eq.(\ref{SolSymDif}).

{An alternative way} to obtain Eq.(\ref{I_cDIF}) is to solve the Usadel
equation, {which was done in many publications }\cite%
{Buzdin91,BuzdinRMP,Pistolesi,Faure}.

For an arbitrary value of the parameter $h\tau_t $ the critical current can
be computed from Eq. (\ref{I_c}). In Fig. 1 we show the dependence of the 
{absolute value of }$I_{c}$ on the thickness $2d$ of the ferromagnet
for different values of $h\tau_t $ taking into account only the scattering
by non-magnetic impurities, i.e when $\tau_{Mt}^{-1}\rightarrow 0$. One can
readily {observe} the crossover from the diffusive case ($h\tau_t =0.2$), {%
when} the decay {length }and {the period of the }oscillations are the same $%
\sqrt{D/h},$ to the quasiballistic case ($h\tau_t =2.2$), {when }the period
of the oscillations is $\pi v_{F}/h$ while the decay length {is of the order
of} the mean free path $l$.

To simplify numerical calculations, we assume that the transmission
parameter $\gamma (\mu )$ is peaked at $\mu =1$ and replace it in Eq.(\ref%
{I_c}) by a delta-function, i.e. we assume that only electrons with momentum
direction perpendicular to the S/F interface are transmitted (the
angle-dependence of ''the transmission coefficient''\ $\gamma (\mu )$
depends on properties of the S/F interface). Another limit ($\gamma (\mu
)=const$) was assumed in Refs.\cite{Blum,Tsukernik} where the expression for 
$I_{c}$ derived in Ref. \cite{BVEpr01} in the absence of depairing
mechanisms was used for comparison between theory and experimental data. 
\begin{figure}[tbp]
\includegraphics[scale=1]{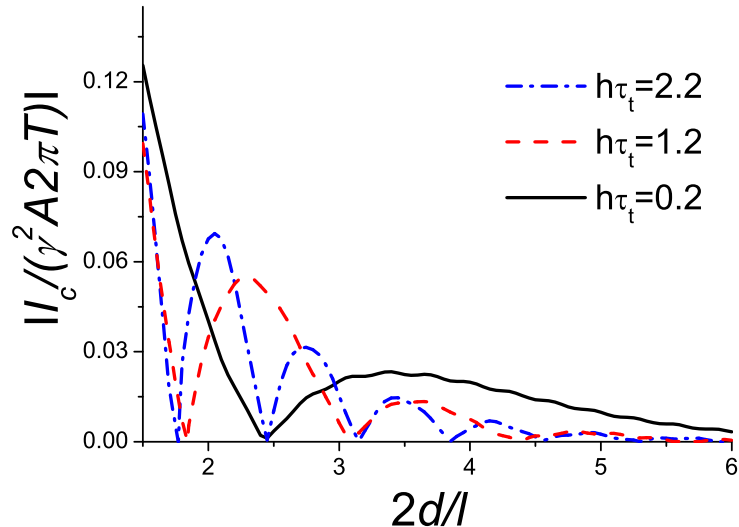}
\caption{Dependence of {the absolute value of the }Josephson critical
current $I_{c}$ on the thickness $2d$ of the ferromagnetic layer for
different values of $h\protect\tau $. Here $\Delta \protect\tau =0.1$ and $T%
\protect\tau =0.05$. Only scattering by non-magnetic impurities is
considered.}
\label{fig1}
\end{figure}
In Fig. 2 we represent the critical current as a function of $2d$ for three
different values of the parameter $h\tau_t$. One can see that in all cases
an increase of $\lambda _{z}$ leads to a decrease of the amplitude of the
condensate in the ferromagnet. 
\begin{figure}[tbp]
\includegraphics[scale=.8]{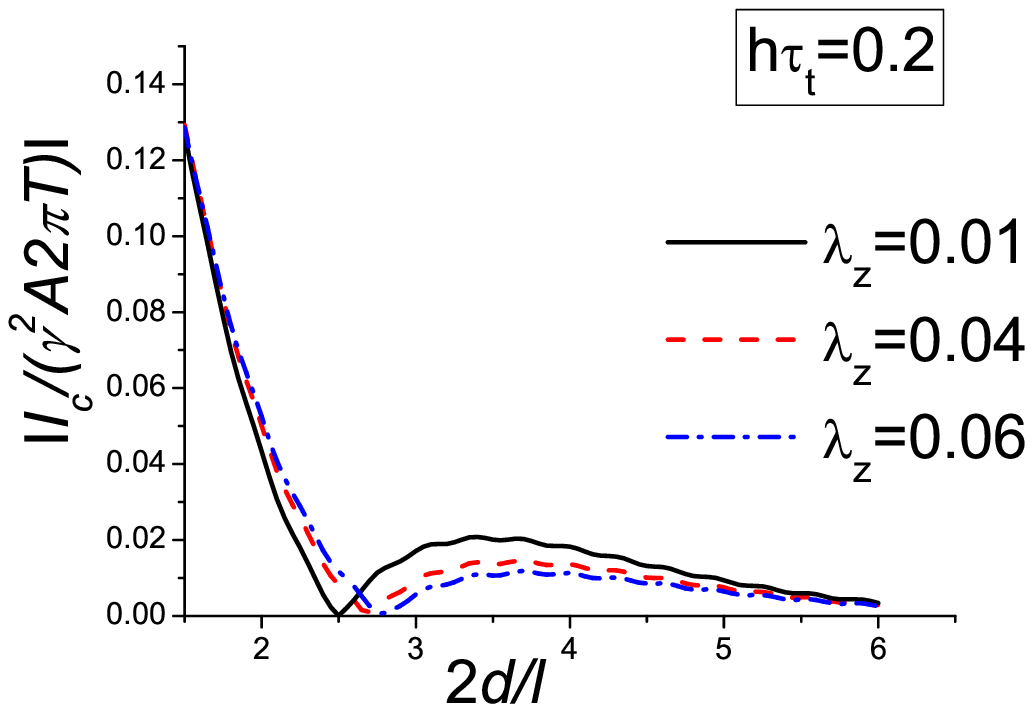} \includegraphics[scale=.8]{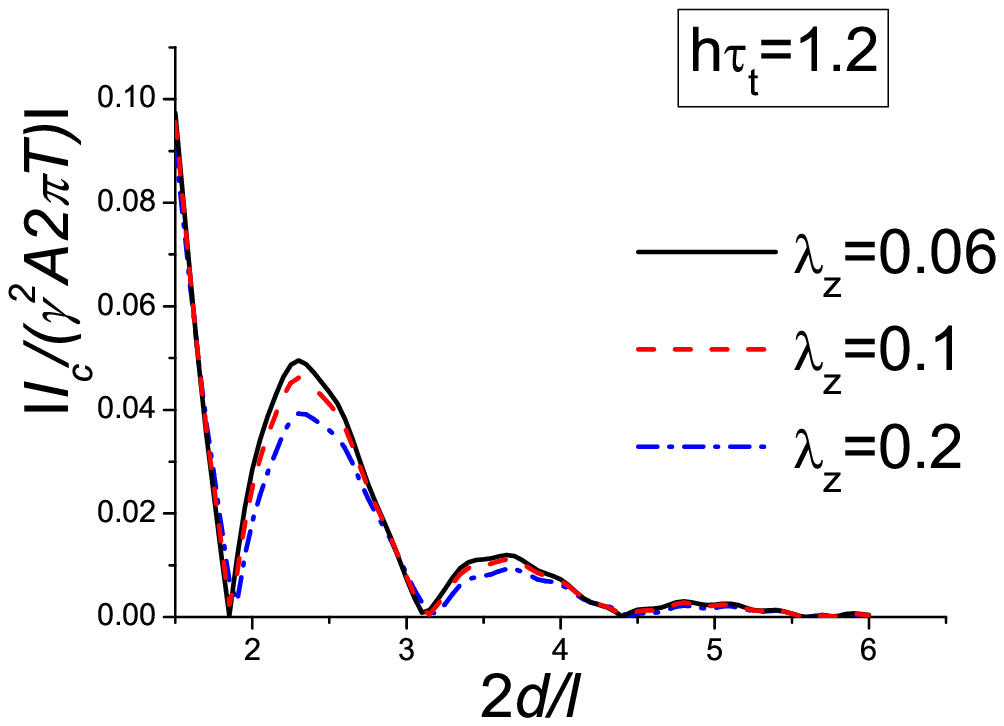} %
\includegraphics[scale=.8]{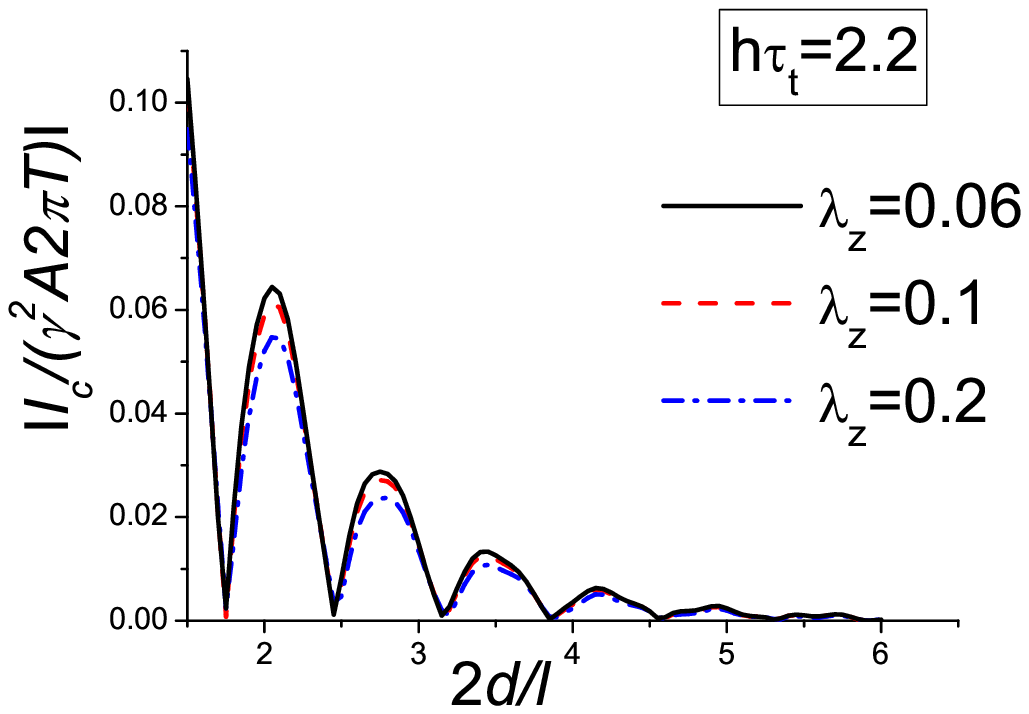}\label{fig2}
\caption{Dependence of {\ the absolute value of } Josephson critical
current $I_{c}$ on the thickness $2d$ of the ferromagnetic layer for
different values of $h\protect\tau $ and $\protect\lambda _{z}$. Here $%
\Delta \protect\tau _{t}=0.1$, $T\protect\tau _{t}=0.05$ and $\protect%
\lambda _{x}=0$. Without loss of generality we have set $\protect\tau%
_N^{-1}\rightarrow0$.}
\end{figure}

Finally we {write down} the expression for the critical current {for the
case }when perpendicular fluctuations of the exchange field are taken into
account, i.e., when $\lambda _{\perp }$ is finite. In that case we obtain
for the critical current a {rather cumbersome formula}
\begin{eqnarray}
I_{c} &=&A(2\pi T)\mathrm{Re}\sum_{n=-\infty,\omega\geq0}^{\infty
}f_{s}^{2}(-1)^n\langle \mu \gamma (\mu )\frac{l}{d}\frac{2\kappa _{+}}{ M_{n}^+(\mu
)\left( |N_{n}^+|^{2}-\lambda _{\perp }^{2}|\kappa _{+}|^{2}\left| \langle
(M_{n}^+)^{-1}\rangle \right| ^{2}\right) }\times  \nonumber \\
&&\times \left[ \left( 1-2\lambda _{z}-\lambda _{\perp }\right) \left(
\kappa _{+}N_{n}^-\left\langle \frac{\mu ^{\prime }\gamma (\mu ^{\prime })}{%
M_{n}^+(\mu ^{\prime })}\right\rangle _{\mu ^{\prime }}-\lambda _{\perp
}|\kappa _{+}|^{2}\left\langle (M_n^{+})^{-1}(\mu ^{\prime })\right\rangle
_{\mu ^{\prime }}\left\langle \mu ^{\prime }\gamma (\mu ^{\prime }){%
(M_{n}^-)^{-1}(\mu ^{\prime })}\right\rangle _{\mu ^{\prime }}\right)
\right. -  \nonumber \\
&&\left. -\lambda _{\perp }\left( \kappa _{-}N_n^{+}\left\langle \frac{\mu
^{\prime }\gamma (\mu ^{\prime })}{M_n^{-}(\mu ^{\prime })}\right\rangle
_{\mu ^{\prime }}-\lambda _{\perp }|\kappa _{+}|^{2}\left\langle
(M_n^{-})^{-1}(\mu ^{\prime })\right\rangle _{\mu ^{\prime }}\left\langle
\mu ^{\prime }\gamma (\mu ^{\prime }){(M_n^{+})^{-1}(\mu ^{\prime })}%
\right\rangle _{\mu ^{\prime }}\right) \right] +\frac{l}{d}\frac{2\kappa _{+}}{M_n^{+}}%
\mu \gamma (\mu )\rangle _{\mu }  \label{I_c2}
\end{eqnarray}
This equation can be evaluated only numerically. We represent the function $%
I_{c}(d)$ for different values of $\lambda _{\perp }$ in Fig. \ref{fig3}.
Note that, again, with increasing $\lambda _{\perp }$ the amplitude of the
condensate decreases. This {is }clear from the physical point of view
because any depairing factors lead to a suppression of the condensate
amplitude. 
\begin{figure}[tbp]
\includegraphics[scale=1]{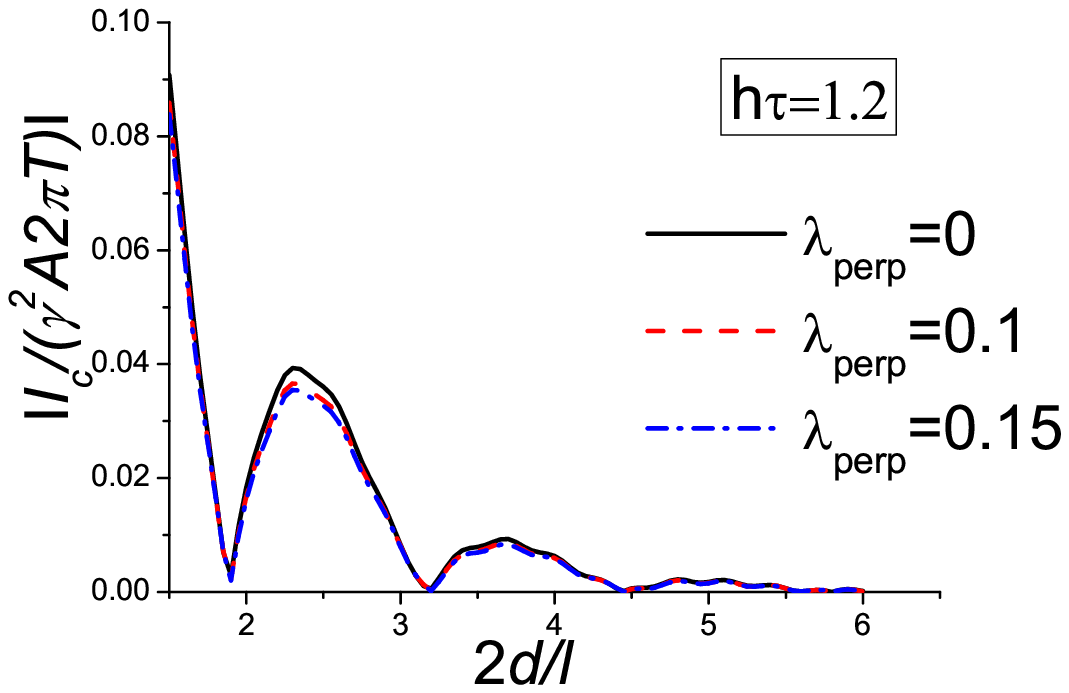}
\caption{Dependence of the {\ the absolute value of } Josephson
critical current $I_{c}$ on the thickness $2d$ of the ferromagnetic layer
for $h\protect\tau =1.2$ and different values of $\protect\lambda _{\perp }$%
. Other parameters are taken as in Fig. \ref{fig1}. }
\label{fig3}
\end{figure}

\section{Spin-orbit scattering}

{In this section we consider} influence of the {spin-orbit (}SO) scattering
on the Josephson current. For simplicity we neglect the spin-dependent
scattering analyzed in the preceding Section. In {the presence of }the SO
scattering, the condensate function is {no longer} diagonal in the spin
space. Therefore, instead of the $2\times 2$ matrix $\hat{f}$, we have to
introduce a more complicated $4\times 4$ matrix $\check{f}$ (see for example %
\cite{BVErmp}).

The Josephson current in SFS junctions in the presence of the SO interaction
was analyzed in the diffusive limit in Ref. \cite{Demler}. Here we focus on
the {opposite }quasiballistic limit.

The linearized Eilenberger equation for $\check{f}$ \ has the form 
\begin{equation}
\mathrm{sgn}\omega \hat{\tau}_{3}l\mathbf{e}\nabla \check{f}+\hat{\kappa}%
\check{f}=\rho \langle \check{f}\rangle -\lambda _{so}\langle \check{f}%
\rangle _{so},  \label{EilSO}
\end{equation}%
{In Eq. (\ref{EilSO}),} $l=v_{F}\tau _{t},\tau _{t}^{-1}=\tau
^{-1}+(2/3)\tau _{so}^{-1},$ {where }$\tau $ and $\tau _{so}$ are the
momentum relaxation time due to a potential scattering and the spin orbit
scattering time, $\rho =\tau _{t}/\tau \approx 1${, }$\lambda _{so}=\tau
_{t}/\tau _{so}<<1$. The matrix $\hat{\kappa}=\kappa (h)=1+2(|\omega
|+ih_{\omega }\hat{\sigma}_{3})\tau _{t}$, {which means} that $\kappa_+$ in
Eq.(\ref{Eil}) is the $(\hat{\kappa})_{11}$ element of the matrix $\hat{%
\kappa}$. The angle brackets mean the angle averaging: 
\begin{equation}
\langle \check{f}\rangle =(1/4\pi )\int d\Omega \check{f}(\Omega ),\langle 
\check{f}\rangle _{so}=(1/4\pi )\int d\Omega ^{\prime }e_{i}^{\prime
}e_{k}^{\prime }(\mathbf{\check{S}\times e})_{i}\check{f}(\Omega ^{\prime })(%
\mathbf{\check{S}\times e})_{k},  \label{AverageSO}
\end{equation}%
where the vector $\mathbf{\check{S}}$ has components $S_{i}$: $\mathbf{%
\check{S}=(}\hat{\sigma}_{1},\hat{\sigma}_{2},\hat{\sigma}_{3}\otimes \hat{%
\tau}_{3}\mathbf{).}$

As before, we represent $\check{f}$ as a sum of the antisymmetric and
symmetric parts: $\check{f}=\check{a}+\check{s}$, where the antisymmetric
part $\check{a}$ is expressed in terms of the symmetric function as $\check{a%
}=-\hat{\kappa}^{-1}\mathrm{sgn}\omega \hat{\tau}_{3}e_{1}l\partial \check{s}%
/\partial x$ and the symmetric part obeys the equation

\begin{equation}
-e_{1}^{2}l^{2}\partial ^{2}\check{s}/x^{2}+\hat{\kappa}^{2}\check{s}=\hat{%
\kappa}[\rho \langle \check{f}\rangle -\lambda _{so}\langle \check{f}\rangle
_{so}+2(e_{1}l)\gamma \sum_{n=-\infty }^{\infty }\check{f}_{s,n}\delta
(x-d(2n+1))]\;,  \label{EqSymSO}
\end{equation}%
{Taking into account the structure of the }$\hat{f}_{s}${\ matrix in the
spin space we see that }the matrix $\check{f}_{s,n}$ coincides with the one
presented above, $\check{f}_{s,n}=\hat{\sigma}_{3}\otimes \hat{f}_{s,n}$.

Our task {now }is to solve Eq.(\ref{EqSymSO}). The presence of the term of
the SO scattering makes this task more difficult {than previously}. In order
to simplify the problem, we use the usual smallness of $\lambda _{so}$ \cite%
{AG}. {An additional simplification comes from using }the quasiballistic
case when the value of $|\kappa |$ is large.

{In order to find the solution of Eq. (\ref{EqSymSO}) }we represent the
Fourier transform $\check{s}_{k}$ in a form of an expansion in the small
parameter $\lambda _{so}$: $\check{s}_{k}=\check{S}_{k}+\delta \check{S}_{k}$%
.

In zero order approximation in $\lambda _{so}$ and in the main approximation
in the parameter $|\kappa |$ {we obtain} 
\begin{equation}
\check{S}_{k}=2\hat{\kappa}le_{1}\gamma \hat{M}_{k}^{-1}(e_{1})\sum_{n=-%
\infty }^{\infty }\check{f}_{s,n}\exp (ikd(2n+1))\;,  \label{S(k)SO}
\end{equation}%
{where  $\hat{M}_{k}(e_{1})$ }is a 2$\times $2-matrix in
spin-space with component (1,1) equals to $M_{k+}(e_{1})$ and component (2,2)%
{\ to $M_{k+}(e_{1})$. } {Although }the matrix $\check{S}_{k}$ is
diagonal in the spin space, the correction $\delta \check{S}_{k}$ is not. It
is equal to

\begin{equation}
\delta \check{S}_{k}=-\hat{\kappa}\hat{M}_{k}^{-1}(\Omega )\lambda
_{so}\langle \check{S}_{k}(\Omega ^{\prime })\rangle _{so}
\label{Delta S(k)}
\end{equation}

Using Eqs.(\ref{S(k)SO}), (\ref{AverageSO}), one can represent the average $%
\langle \check{S}_{k}(\Omega ^{\prime })\rangle _{so}$ as%
\begin{equation}
\langle \check{S}_{k}(\Omega ^{\prime })\rangle _{so}=\langle \hat{S}%
_{k0}(\Omega ^{\prime })\otimes \hat{\sigma}_{0}A_{0}-\hat{S}_{k3}(\Omega
^{\prime })\otimes \hat{\sigma}_{3}A_{3}+2\hat{\tau}_{3}\hat{S}_{k0}(\Omega
^{\prime })\otimes i(\hat{\sigma}_{1}A_{1}-\hat{\sigma}_{2}A_{2})\rangle
\label{AverageSO1}
\end{equation}%
where $A_{0}=A_{0}(\Omega ,\Omega ^{\prime })=e_{1}^{\prime
2}(e_{3}^{2}-e_{2}^{2})+e_{2}^{\prime 2}(e_{3}^{2}-e_{1}^{2})+e_{3}^{\prime
2}(e_{1}^{2}+e_{2}^{2});$ $A_{3}=A_{3}(\Omega ,\Omega ^{\prime
})=e_{1}^{\prime 2}(e_{3}^{2}+e_{2}^{2})+e_{2}^{\prime
2}(e_{3}^{2}+e_{1}^{2})+e_{3}^{\prime 2}(e_{1}^{2}+e_{2}^{2});$ $%
A_{1}=e_{1}^{\prime 2}e_{2}e_{3};A_{2}=e_{2}^{\prime 2}e_{1}e_{3}$. The
matrices $\hat{S}_{k0,3}$ are defined {with the help of the relation} $%
\check{S}_{k}=\hat{\sigma}_{0}\otimes \hat{S}_{k0}+\hat{\sigma}_{3}\otimes 
\hat{S}_{k3}$.

{Using} Eq.(\ref{S(k)SO}) we {obtain} 
\begin{equation}
\hat{S}_{k0}=2le_{1}\gamma i\hat{F}_{s}\mathrm{Im}\frac{\kappa_+ }{M_{k+}};%
\hat{S}_{k3}=2le_{1}\gamma \hat{F}_{s}\mathrm{Re}\frac{\kappa_+ }{M _{k+}}
\label{S(k)SO2}
\end{equation}%
where $\hat{F}_{s}=\sum_{n=-\infty }^{\infty }\hat{f}_{s,n}\exp (ikd(2n+1))$
and the matrix $\hat{f}_{s,n}$ {has been introduced} in Eq.(\ref{S(x)}).

Note that due to the last two terms in Eq.(\ref{AverageSO1}) the condensate
matrix $\delta \check{S}_{k}$ contains not only {the }triplet component with 
{the }zero projection {on} the $z$-axis but also {the }triplet components of
the type $\uparrow \uparrow ,\downarrow \downarrow $. However, in the lowest 
{order} in $\lambda _{so}$ these components do not contribute to the
Josephson current because they are odd functions with respect to the
inversion $e_{1,2}\Rightarrow -e_{1,2}$.

The correction $\delta I_{c}$ to the Josephson current due to spin orbit
scattering {is given by the expression} 
\begin{equation}
\delta I_{c}=A(2\pi iT)Tr\left[ \hat{\tau}_{3}\otimes \hat{\sigma}%
_{0}\sum_{\omega =-\infty }^{\infty }\int \frac{dk}{2\pi }\left\langle
e_{1}\gamma \lbrack \delta \check{S}_{k}(\Omega ),\check{f}%
_{s}(d)]_{-}\right\rangle \right] \exp (-ikd),\;  \label{JosCurSO}
\end{equation}


Finally, we obtain the correction $\delta I_{c}$ to the critical current
originating from the SO scattering 
\begin{equation}
\delta I_{c}=A(2\pi T)\frac{l}{d}\sum_{n,\omega =-\infty }^{\infty
}f_{s}^{2}\left\langle \langle e_{1}e_{1}^{\prime }\gamma (\Omega )\gamma
(\Omega ^{\prime })[-A_{0}\mathrm{Im}\frac{\kappa _{+}}{M_{2n}^{+}(\Omega )}%
\mathrm{Im}\frac{\kappa _{+}}{M_{2n}^{+}(\Omega ^{\prime })}+A_{3}\mathrm{Re}%
\frac{\kappa _{+}}{M_{2n}^{+}(\Omega )}\mathrm{Re}\frac{\kappa _{+}}{%
M_{2n}^{+}(\Omega ^{\prime })}]\rangle _{\Omega ^{\prime }}\right\rangle
_{\Omega }\;  \label{JosCurSO2}
\end{equation}%
The structure of this equation is similar to the {one} of the contribution
of the first term in the square brackets in Eq.(\ref{I_c}) to the Josephson
current. This contribution is also small in comparison with the current $%
I_{c}$ determined by Eq.(\ref{I_cQB}) with $L_{qb}(\mu )=l|\mu |/\kappa $
and $l=v_{F}\tau _{t}$, where $\tau _{t}^{-1}=\tau ^{-1}+(2/3)\tau
_{so}^{-1} $. This small correction to the critical current $I_{c\text{ }}$%
can be calculated numerically. In the diffusive case the SO scattering has
been studied in Refs.\cite{Buzdin91,BuzdinRMP,Pistolesi,Faure}.

\section{Conclusions}

Assuming a weak proximity effect we have derived the exact expression for
the Josephson current through an SFS junction for arbitrary impurity
concentration and {in the presence of }spin-dependent scattering. In the
quasiballistic and diffusive limits this expression {takes a simple form}.
In the {former} case, the parameter $1/(h\tau )$ is small. In the main
approximation the expression for the critical Josephson current is reduced
to Eq.(\ref{I_cQB}) {that agrees} with the corresponding equation {of} Ref.%
\cite{BVEpr01} {provided} the momentum relaxation time $\tau $ is replaced {%
as}: $\tau ^{-1}\Rightarrow \tau _{t}^{-1}=(1+\lambda _{z})\tau
^{-1}+(2/3)\tau _{so}^{-1}$. Therefore, the deparing leads only to a
renormalization of the mean free path $l$ {that}\ determines {the }decay of
the condensate function in F and of the Josephson critical current $I_{c}$.
Oscillations of these quantities have the period $\pi v_{F}/h$.

In the diffusive case the oscillation period and {the} decay of the critical
current $I_{c}$ are determined by the value of the product $h\tau _{dep}$
where the time $\tau _{dep}$ is defined as $\tau _{dep}=\min \{\tau
_{sp},\tau _{so}\}$. If the exchange energy lies in the interval $\tau
_{dep}^{-1}<h<\tau ^{-1}$, then the period of the oscillations of $I_{c}$ is 
$2\pi \sqrt{D\tau _{dep}}/(h\tau _{dep})$ and the decay length is $\sqrt{%
D\tau _{dep}}$. In the limit $h<\tau _{dep}^{-1}$ the period of oscillations
and the decay length are determined by the value $\sqrt{D/h}$. The general
expression, Eq.(\ref{I_c}), may serve for numerical calculation of the
critical current in the intermediate {region of parameters.}

\section*{Acknowledgment}

We would like to thank M. Kharitonov for useful discussions and comments.
This work was supported by SFB 491. A.F.V. thanks the DFG for finincial
support within {the program} \textquotedblleft
Mercator-Gastprofessoren\textquotedblright .  F.S.B. acknowledges funding by the Ram\'on y Cajal program.

\end{document}